# A critical evaluation of dynamic fracture simulations using cohesive surfaces

Michael L. Falk, Alan Needleman, James R. Rice

**Abstract**. Finite element calculations of dynamic fracture based on embedding cohesive surfaces in a continuum indicate that the predictions are sensitive to the cohesive law used. Simulations were performed on a square block in plane strain with an initial edge crack loaded at a constant rate of strain. Cohesive laws that have an initial elastic response were observed to produce spontaneous branching at high velocity, but to modify the linear elastic properties of the body. As a consequence the cohesive surface spacing cannot be refined arbitrarily and becomes an important length scale in the simulations. Cohesive laws that are initially rigid do not alter the linear elastic response of the body. However, crack branching behavior was not observed when such a cohesive relation was implemented using a regular finite element mesh.

## 1. INTRODUCTION

The phenomenon of dynamic fracture is of interest in a wide variety of contexts from high-speed manufacturing and defense applications to geophysics and seismology. Experimental investigations of dynamic fracture clearly demonstrate the difficulties involved in the predictive modeling of fast moving cracks. Observation has shown that in a wide variety of isotropic homogeneous materials high-speed brittle cracks never reach the theoretically predicted limiting speed, the Rayleigh wave speed. Ravi-Chandar and Knauss first associated the low limiting speed with intermittent propagation, branching and micro-cracking in the vicinity of the crack tip [1]. Recent work by Fineberg, Gross, Sharon and others has indicated that this may be due to a branching instability [2-4]. Since cracks which are moving sufficiently rapidly can always supply enough energy to drive two fractures by slowing down, branching cannot be precluded on energetic grounds. Within the framework of classical fracture mechanics, a branching criterion needs to be specified explicitly. A cohesive surface methodology as in [5-6], which stems from the formulation originally proposed by Dugdale and Barenblatt [7-8], provides a framework for directly simulating crack branching phenomena. However, the extent to which the predictions of this methodology depend on the details of the cohesive constitutive description remains to be determined.

Several cohesive surface formulations have been used to simulate dynamic fracture in brittle solids [5-6]. In [5], the cohesive constitutive relation was taken to be initially elastic, while in [6] the cohesive constitutive relation was taken to be initially rigid. The sensitivity of crack branching predictions to this aspect of the cohesive description is investigated here. It also is of interest to determine whether these methods are interchangeable and to what extent their predictions depend on the particulars of their implementation including such non-physical details as the refinement of the mesh.

Lattice models provide another approach to analyzing fracture instabilities and exhibit branching attempts which may be related to those observed in experiment [9]. A good deal of work, however, remains to be done in relating these methods to a more general continuum mechanics context. In particular, the development of rigorous methods for the simulation and prediction of materials' failure in non-crystalline solids, or any solid where the crack tip process zone plays a significant role, would lie outside the purview of such methods in their current form. In addition it remains of interest to understand the crack branching phenomenon at scales much larger than the atomistic scale.

The dynamic propagation of a crack through a square block in plane strain with a small initial edge crack is analyzed. The calculations are carried out within a finite deformation framework with an elastic constitutive relation for the material and employing several cohesive constitutive relations. Numerically, the governing equations are solved using a finite element method. One of the cohesive constitutive relations is an elastic relation as implemented in Xu and Needleman [5]. Because the stiffness of the block depends on the cohesive properties as well as on the properties of the volumetric constitutive relation, wave speeds depend on the cohesive surface spacing. The other cohesive constitutive relation is similar to the one initiated by Camacho and Ortiz [6] except for its response during unloading. The initial cohesive response is rigid and there is a finite traction at which separation initiates. For this cohesive constitutive relation, there are potential cohesive surfaces but these are not activated until a critical traction is attained. Hence, prior to attaining this critical traction, wave speeds are not affected. Calculations are carried out comparing the crack branching predictions of these two cohesive characterizations.

In some of the calculations here, as in [5-6], cohesive surfaces (or potential cohesive surfaces) were introduced along all boundaries in a finite element mesh where cracks may initiate and propagate. When cohesive surfaces are introduced along all finite element boundaries, the mechanical response clearly depends on the mesh spacing as well as on physical parameters. We also carry out simulations in which the spacing of cohesive surfaces is independent of the mesh size. This introduces another physically meaningful length scale into the problem. However, even in this limit the two methods result in different crack branching behavior. The cohesive constitutive relation with an initial elastic response exhibits crack branching at crack speeds below the Rayleigh wave speed, while the cohesive constitutive relation with an initially rigid response does not.

## 2. METHODOLOGY

The principle of virtual work is expressed as

$$\int_V \mathbf{S} : \delta \mathbf{E} \, dV + \int_{\Gamma_{int}} \mathbf{T} \cdot \delta \mathbf{\Delta} \, d\Gamma = \int_{\Gamma_{ext}} \mathbf{T} \cdot \delta \mathbf{u} \, d\Gamma - \int_V \rho \ddot{\mathbf{u}} \cdot \delta \mathbf{u} \, dV, \tag{2.1}$$

where the integrals are taken over the volume, external surface area and internal cohesive surface area of the body in the reference configuration, see e.g. [5]. $\mathbf{\Delta}$ is the displacement jump across the cohesive surface, and the work-conjugate traction vector on that surface is denoted $\mathbf{T}$. The dots over $\mathbf{u}$ denote derivatives with respect to time. Following the analysis in [5] elastic constants consistent with PMMA have been chosen: $E = 3.24$ GPa, $\nu = 0.35$ and $\rho = 1.19$ g/cm$^3$. This results in dilational, shear and Rayleigh wave speeds of $c_d$=2090 m/s, $c_s$=1004 m/s and $c_R$=938 m/s. The constitutive law for the cohesive surfaces, a relation between the traction and the separation along the internal surfaces, is expressed in terms of a potential function, $\varphi$, such that the traction $\mathbf{T} = \partial \varphi / \partial \mathbf{\Delta}$. The specific form of $\varphi$ will depend on the particular model in question. In general, so long as $\varphi$ is history and rate independent the cohesive surfaces are fully reversible and the fracture energy is well defined and independent of crack speed.

### 2.1. Cohesive surfaces with initial linear response

In all the models discussed here traction across the cohesive surface is a continuous function of a displacement jump. In the class of models we will refer to as initially linear models, for a purely normal displacement jump, the magnitude of the normal traction increases monotonically from zero for a compressive displacement jump and, in tension, the magnitude of the traction initially increases from zero, reaches a maximum and then decreases to zero. For tangential displacement jumps, the shear traction increases to a maximum and decays to zero regardless of sign. This shear behavior is an empirical model of the nonlinear mode II fracture response and not of inter-atomic slip as in the Peierls model. Unless otherwise specified, as in [5], all finite element boundaries are taken to define cohesive surfaces. Due to the finite stiffness at small values of the displacement jump across the cohesive surface, the stiffness of the cohesive surfaces contributes to the linear elastic response of the solid, a feature that will prove important for understanding the resulting fracture behavior. In the cohesive constitutive relation of [5] the potential is of the form

$$\varphi_{XN}(\mathbf{\Delta}) = -\sigma_{max}(\delta_0 + \Delta_n)\exp\left(1 - \frac{\Delta_n}{\delta_0} - \frac{\Delta_t^2}{\delta_0^2}\right) \tag{2.2}$$

Here $\sigma_{max}$ is the maximum normal traction. The maximum traction in shear is equal to $\tau_{max} = \sqrt{2e}\sigma_{max} \approx 2.3\sigma_{max}$. Figure 1 shows the resulting separation dependent traction,

$$T_n = \sigma_{max}\left(\frac{\Delta_n}{\delta_0}\right)\exp\left(1 - \frac{\Delta_n}{\delta_0} - \frac{\Delta_t^2}{\delta_0^2}\right) \qquad T_t = 2\sigma_{max}\frac{\Delta_t}{\delta_0}\left(1 + \frac{\Delta_n}{\delta_0}\right)\exp\left(1 - \frac{\Delta_n}{\delta_0} - \frac{\Delta_t^2}{\delta_0^2}\right) \tag{2.3}$$

The maximum traction together with the length scale $\delta_0$ determines the surface energy $\gamma = \delta_0 e \sigma_{max}/2$ In keeping with the example in [5] a fracture toughness appropriate for PMMA is chosen: $\gamma = 176.15$ J/m$^2$. This value is too large for the actual fracture process to be truly elastic, so our potential formulation is a proxy for an irreversible separation.

## 2.2. Cohesive surfaces with initial rigid response

The above formulation has the feature that the cohesive surfaces contribute to the linear elastic response of the body, which poses limitations that will be discussed in more detail in section 3.2. We can, however, consider a cohesive law implementation that does not include such a feature. Camacho and Ortiz have studied one such implementation, albeit with the additional feature of irreversibility [6].

Implementation of a version of the cohesive surface model without linear response would, at its simplest, seem to require a potential of the form

$$\varphi_{CO}(|\mathbf{\Delta}|) = \sigma_{max}\left(|\mathbf{\Delta}| - \frac{|\mathbf{\Delta}|^2}{4e\delta_0} - e\delta_0\right)\Theta(2e\delta_0 - |\mathbf{\Delta}|) \tag{2.4}$$

$$|\mathbf{\Delta}| = \sqrt{\Delta_n^2 + \beta_t \Delta_t^2}$$

where $\beta_t$ is a parameter that determines the ratio of maximum shear traction to maximum normal traction; $\Theta$ is the Heaviside step function. This is equivalent to the cohesive surface implementation of Camacho and Ortiz [6] without irreversibility. Note that, unlike the potential in Xu and Needleman [5], Eq. (2.2), this potential is only valid when $\Delta_n$ is positive and the derivative of $\varphi$ is discontinuous when $\Delta_n$ crosses zero since negative values of $\Delta_n$ correspond to overlap between rigid elements. In this algorithm contact forces between elements prevent overlap. The approximation used to handle contact will be discussed further at the end of this section. Figure 1 shows the tractions that result from this potential,

$$T_n = \sigma_{max}\left(\frac{\Delta_n}{|\Delta|} - \frac{\Delta_n}{2e\delta_0}\right)\Theta(2e\delta_0 - |\Delta|), \quad T_t = \sigma_{max}\left(\frac{\beta_t \Delta_t}{|\Delta|} - \frac{\beta_t \Delta_t}{2e\delta_0}\right)\Theta(2e\delta_0 - |\Delta|) \quad (2.5)$$

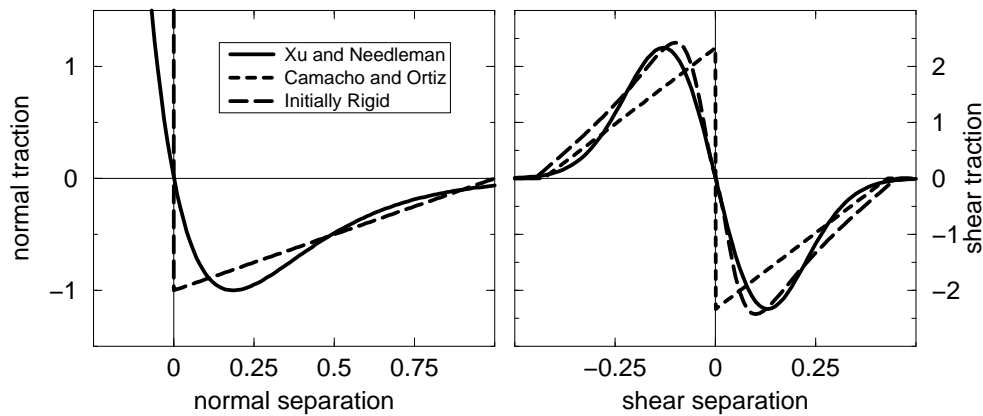

**Figure 1.** The traction versus separation curves for the Xu-Needleman model, the Camacho-Ortiz model (in the case of no unloading) and the initially rigid model implemented here. The tractions are given in units of the maximum normal stress, $\sigma_{max}$, and the displacements are given in units of the maximum normal displacement $2e\delta_0$. Under normal separation the Camacho-Ortiz and initially rigid models are identical.

It is apparent from Eqs.(2.5) that the maximum shear traction is related to the maximum normal traction by $\tau_{max} = \sqrt{\beta_t}\sigma_{max}$, and the cohesive opening criterion can be written

$$\mathcal{F}(\mathbf{T}) = T_n^2 + T_t^2/\beta_t - \sigma_{max}^2 = 0 \quad (2.6)$$

The maximum normal distance before complete separation is $2e\delta_0$.

In the above cohesive surface formulation $T_t$ is discontinuous around $\Delta_t \approx 0$ when $\Delta_n = 0$. This discontinuity leads to a numerical instability when the cohesive surface experiences a small shear perturbation after breaking. To eliminate this problem and preserve the reversible character of the cohesive surfaces a second length scale $\delta_\varepsilon$ is introduced in the shear direction below which the cohesive surface behaves in a linear manner after separation. Note that this initial linear response only occurs once the traction on the surface exceeds a critical traction after which the cohesive surface is considered to have split. This model will be referred to as the initially rigid model, and it consists of a potential of the form

$$\varphi_R = \sigma_{max}\frac{\left(|\Delta| - \frac{|\Delta|^2}{4e\delta_0} + \delta_\varepsilon\right)\Delta_t^2 + \left(\Delta_n - \frac{\Delta_n^2}{4e\delta_0}\right)\delta_\varepsilon^2}{\Delta_t^2 + \delta_\varepsilon^2} - \sigma_{max}e\delta_0 \quad (2.7)$$

This potential is equivalent to $\varphi_{CO}$ when $\Delta_t=0$. When $|\Delta_t| \gg \delta_\varepsilon$ it reduces to $\approx \varphi_{CO} + \sigma_{max}\delta_\varepsilon$. But, when $|\Delta_t| \ll \delta_\varepsilon$ the potential becomes

$$\varphi_R \approx \sigma_{max}\left(\frac{\Delta_t^2}{\delta_\varepsilon} + \Delta_n - \frac{\Delta_n^2}{4e\delta_0} - e\delta_0\right) \quad (2.8)$$

The tractions resulting from this cohesive law are illustrated in Figure 1 for $\delta_\varepsilon = 2\,e\,\delta_0/10$.

In the cohesive surface implementation with initially rigid cohesive surfaces although all the element corners have an associated node many of these nodes are associated with a common vertex and do not move independently. We will refer to such nodes as being "bound." Each set of bound nodes must be checked regularly to determine if it will "split" along any of its potential cohesive surfaces. Once split these nodes can move independently though they may interact with each other via the cohesive interaction along the now split surface. To determine when splitting will occur the tractions along the bound surfaces must be computed and the cohesive opening condition, Eq. (2.6), tested. The surface will split if $\mathcal{F}(T) \geq 0$.

Consider first the calculation of the traction along a surface at an entirely bound node. Assuming that no external tractions are acting along the surfaces in question the nodal forces are computed and used to determine the acceleration of the node as determined by Eq. (2.1).

$$\mathbf{F}_j = -\int_V (\mathbf{S}:\mathbf{B}_j)dV = \int_V (\rho\ddot{\mathbf{u}} \cdot \mathbf{N}_j)dV \quad (2.9)$$

Here $\mathbf{B}_j$ is the variation in strain associated with node j and $\mathbf{N}_j$ is the variation in displacement associated with node j. In order to determine the traction along any of the surfaces that cross this node we can compute the nodal forces from elements above and below the surface.

$$\mathbf{F}_j^\pm \equiv -\int_{V\pm}(\mathbf{S}:\mathbf{B}_j)dV \quad (2.10)$$

Here V± denotes the integral restricted to the region above or below the surface in question. Since linear elements are being used the net force across the surface must equal the traction along the surface integrated over the surface area times a linear function. For a surface along the x direction from –L to L

$$\int_{-L}^{L} dx \left(1 - \frac{|x|}{L}\right) \mathbf{T}_j = \mathbf{F}_j^+ - \mathbf{F}_j^-. \tag{2.11}$$

The traction is then computed to be $\mathbf{T}_j = \left(\mathbf{F}_j^+ - \mathbf{F}_j^-\right)/L$ in the case of a previously bound node.

For nodes at which one or more surfaces have already split V+ instead denotes only the region clockwise from the surface in question until a split surface is reached and V- denotes the corresponding region in the anti-clockwise direction. It is then also necessary to include the contributions from the side Γ+ or Γ- of the cohesive surface that bounds the V+ or V- region in the calculation of the nodal forces.

$$\mathbf{F}_j^{\pm} \equiv -\int_{V_{\pm}} (\mathbf{S} : \mathbf{B}_j) dV - \int_{\Gamma_{\pm}} (\mathbf{T} \cdot \mathbf{N}_j) d\Gamma \tag{2.12}$$

where the normal vector $\hat{n}$ on Γ± points into the domain V+ or V-. In the case of a previously split node the traction is only computed along the region from 0 to L and the traction that results is $\mathbf{T}_j = 2\left(\mathbf{F}_j^+ - \mathbf{F}_j^-\right)/L$.

After cohesive surfaces have completely split it is still possible for these surfaces to again make contact or even exert normal repulsive forces on each other if their normal separation becomes zero. It is necessary to check for and handle such contact events. This is done after the nodes are moved each time step. Cohesive surfaces that would have overlapped are assumed to have undergone an inelastic collision instead. To simulate contact the nodes are repositioned at their combined center of mass and are assigned a velocity that preserves the total momentum of the system. Note that energy is not conserved in this process.

## 2.3. Numerical Implementation

We implement the above procedures in plane strain. The region analyzed is a 3mm square block with an edge crack extending 0.25mm into the block from the left edge. The block is loaded uni-axially by constantly increasing the relative y-displacements of the top and bottom boundaries while zero shear traction is maintained along these boundaries. The side boundaries are stress free. The initial conditions consist of uniformly applied velocity gradients in the x and y directions with strain rates $\dot{\varepsilon}_{yy} = 2 \times 10^3$ s$^{-1}$ and $\dot{\varepsilon}_{xx} = 0$. The velocities of the top and bottom boundaries are not altered during the simulation. This loading does not generate waves propagating in from the top and bottom boundaries, although these boundaries will reflect waves generated by the propagating crack tip. The stress intensity factor of the crack increases during the simulation both due to the increasing strain and due to the changing crack length.

The finite element mesh is composed of uniformly sized triangular elements arranged in a "crossed-triangle" pattern described in [5]. The uniformity of the mesh minimizes dispersion effects. The number of degrees of freedom is six times the number of elements. However in the initially rigid formulation, prior to separation, many of the nodes are slaved to each other and do not contribute independent degrees of freedom. The mesh for the square block is composed of 96×96 31.25 μm quadrilaterals, 120×120 25 μm quadrilaterals or 240×240 12.5 μm quadrilaterals.

The equations of motion are integrated using a central-difference explicit direct method formally equivalent to a Newmark β-method with β=0. This method is second order accurate. The time step was chosen to be a fraction of the above mentioned natural time step for the procedure $\Delta_t = 10^{-9}$ s $\approx 0.1$ $l_m/c_d$ where $l_m$ is the mesh size.

## 3. SIMULATIONS

### 3.1. Length Scales

It is important to note four length scales in these simulations. The first of these is a macroscopic scale $L$ that characterizes the geometry of the body, which, for example, may be identified with the crack length or the length of the remaining ligament. The mesh size $l_m$ provides a non-physical length scale, and it is clearly necessary that $L$ is much larger than $l_m$ for the mesh to provide an accurate resolution of the stress near the crack tip. The cohesive zone length $l_z$ is a measure of the length over which the cohesive constitutive relation plays a role. The cohesive zone length is set by the elastic properties of the solid, the maximum traction $\sigma_{max}$ and the surface energy $\gamma$. For the simple finite traction separation potential $\phi_{CO}$ this length at zero velocity is approximately [10]

$$l_z(v=0) = \frac{9\pi}{32}\left(\frac{E}{1-v^2}\right)\frac{2\gamma}{\sigma_{max}^2} \tag{3.1}$$

For a choice of $\sigma_{max} = E/10$ as in the calculations in [5], $l_z(v=0) \approx 11$ μm. This choice of $\sigma_{max}$ has the consequence that $l_m > l_z$ for all the mesh sizes used here and the cohesive zone would clearly not be resolved. Therefore, in the calculations here $\sigma_{max} = E/25$, $l_z \approx 68$ μm and the mesh size is comparable to or smaller than the cohesive zone length. Still there are only 2-5 elements over the cohesive zone length so that variations within the cohesive zone may still not be fully resolved.

### 3.2. Simulations with branching between all elements

There are two competing requirements when the boundaries of all finite elements are taken to be cohesive surfaces and when the cohesive surfaces have an initial linear elastic response: (i) the cohesive contribution to the stiffness should be small compared to that of the volumetric constitutive relation and (ii) the mesh size needs to be less than the cohesive zone length as given by Eq. (3.1). The first requirement is met if the spring constant of the cohesive surfaces, calculated from Eq. (2.2) and the mesh size, is much greater than the Young's modulus of the material.

$$E \ll l_m \sigma_{max} e / \delta_0 = l_m (\sigma_{max} e)^2 / 2\gamma \tag{3.2}$$

This is met when the value of $\sigma_{max}$ is such that the dimensionless group

$$\Xi \equiv \frac{\gamma E}{l_m \sigma_{max}^2} \ll 1. \tag{3.3}$$

To meet the second requirement, however,

$$l_m < l_z = \frac{\zeta \gamma E}{\sigma_{max}^2} \quad \Rightarrow \quad \Xi \gg \frac{1}{\zeta} \tag{3.4}$$

where for a Poisson's ratio of 0.35, $\zeta \approx 2$. These two conditions cannot be satisfied simultaneously so that it is not possible to resolve the cohesive zone without affecting the wave speeds of the solid.

Different issues come to the fore for initially rigid cohesive relations. In the quasi-static limit the largest normal traction in the elastic body is exactly at the tip of the cohesive zone. However, for the moving crack the highest tensile stresses are not confined to the tip and occur in a region around the cohesive zone. Therefore, we expect that at non-zero speed a region of material around the tip may develop multiple cracks. The amount of "damage" in this crack tip region will depend on the resolution of the mesh in the vicinity of the crack tip. Figure 2 shows an example of a crack tip in this regime. The diffuse region of cohesive surface opening is evident in the crack tip region. The development of this damage region seems to coincide with a "wiggly" crack path as evident in Figure 2. This is not identical behavior to the macroscopic branching seen in [5] but rather consists of the crack jumping by one or two elements, a distance comparable to the cohesive zone length, parallel to the initial crack plane.

## 3.3. Simulations with a well defined distance between cohesive surfaces

The problems that arise when cohesive surfaces exist between all elements can be avoided by introducing an independent distance between cohesive surfaces, $l_c >> \Xi\, l_m = l_z/\zeta$. We have run a series of simulations using initially elastic and initially rigid cohesive constitutive laws in which the distance between cohesive surfaces $l_c = 125$ µm (120×120, 240×240) or 187.5 µm, (80×80). The cohesive zone length $l_z$ in Eq. (3.1) is approximately 68 µm, while the three element sizes studied are $l_m = 37.5$ µm (80×80), 25 µm (120×120), 12.5 µm (240×240). Therefore $l_c > l_z > l_m$ for all the simulations presented here.

Figure 3 shows the changes in the crack tip location and crack tip speed versus time in simulations using the two larger mesh sizes for both cohesive laws. In these graphs it is difficult to distinguish between the 120×120 data and the 240×240 data because the simulations produce very similar results. This indicates some level of convergence despite the fact that we are only marginally resolving the cohesive zone length in these simulations. The most striking difference between the simulations with initially elastic and initially rigid cohesive surfaces is that all the initially elastic simulations reach a crack speed approaching the Rayleigh wave speed and then exhibit a drop in crack speed followed by macroscopic branching. The vertical line in the figure denotes the onset of macroscopic branching in the simulations with initially elastic cohesive surfaces. The simulations with initially rigid cohesive surfaces never display the drop in crack speed or the macroscopic branching as evident in Figure 4.

## 4. CONCLUSIONS

The calculations reported on here illustrate that simulations of crack branching with cohesive surfaces between all elements are not well posed in the sense that the results continually change with mesh refinement. Introducing an additional length scale into the problem, the distance between cohesive surfaces, circumvents this problem. The present results suggest that with this additional length scale the prediction of crack branching strongly depends on whether or not the cohesive constitutive relation is initially linear or is initially rigid – crack branching is only obtained in the case of an initially linear response. Preliminary results of Pandolfi and Yu [12] show a similar absence of crack branching in systems with initially rigid response as introduced by Camacho and Ortiz [6]. Nevertheless, since the separation criterion and its numerical implementation are key when the cohesive response is initially rigid, the possibility remains that the lack of branching is a consequence of the numerical implementation. On the other hand, the numerical results are consistent with Gao's [11] contention that the reduction in wave speed due to nonlinear elastic effects plays a major role in precipitating crack branching.

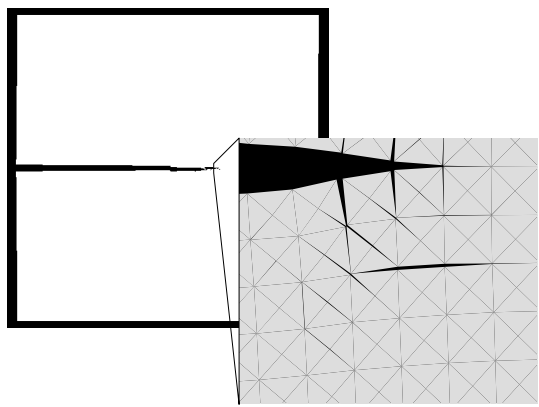
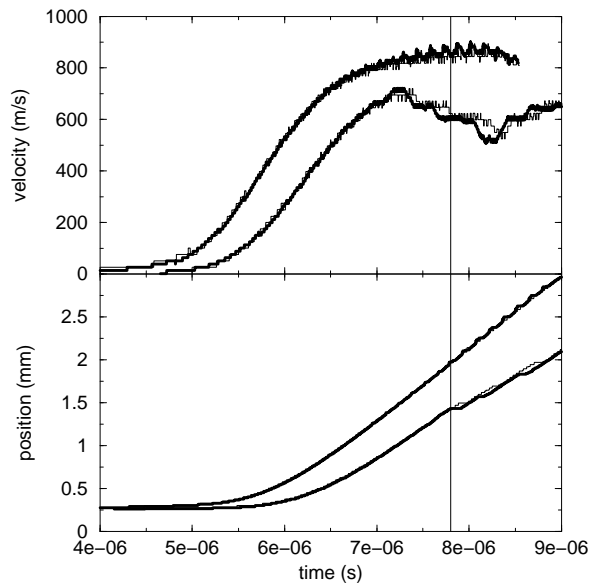

**Figure 2.** A close-up of the tip region in a 240×240 element simulation of a dynamically propagating crack where initially rigid cohesive surfaces are allowed to open between all elements. The cohesive surfaces open in a diffuse region around the crack tip when high speeds are reached. Mesh dependent deviations in the crack path are evident.

**Figure 3.** Crack tip position and velocity vs. time for two cohesive laws. The upper set of lines in each graph corresponds to simulations with initially rigid cohesive surfaces, the lower set to simulations with initially elastic cohesive surfaces. Both a 120×120 (thin line) and a 240×240 element mesh (thick line) were simulated in each case. Simulations with the initially elastic law show a drop in speed followed by branching at the time denoted by the vertical line.

**Acknowledgements**
The authors appreciatively thank A. Pandolfi and R. Yu for sharing their simulation results, and acknowledge M. Ortiz for helpful discussions. MLF and JRR are grateful for the support provided by ONR grant N00014-96-10777 and by the NSF via Harvard University's MRSEC. AN gratefully acknowledges the support of ONR grant N00014-97-1-0179.

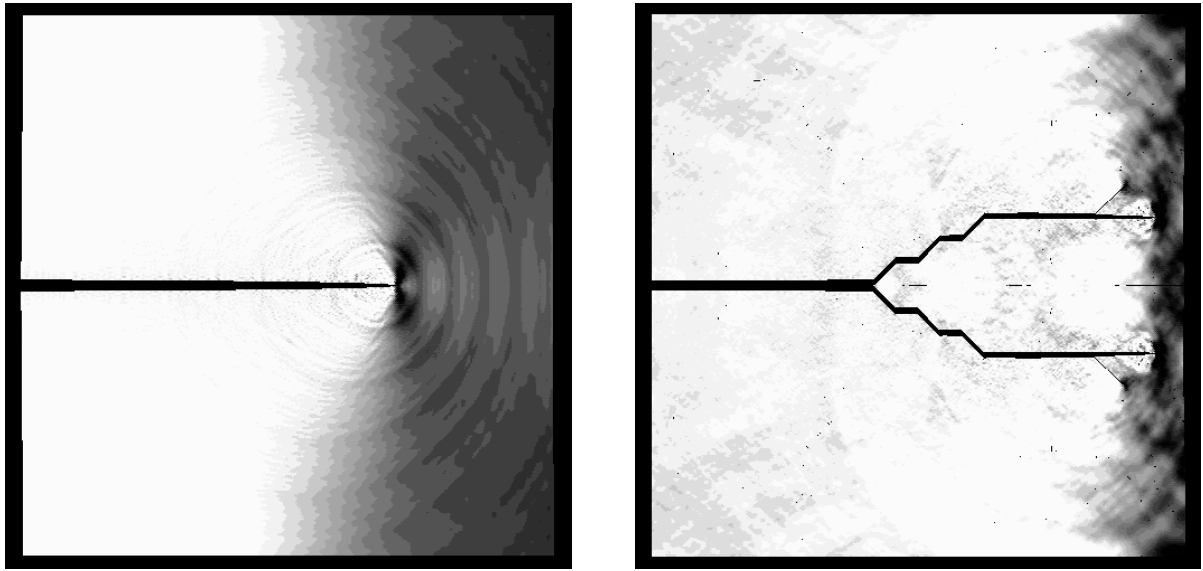

**Figure 4.** The late stages of two 240×240 element simulations of a 3mm square block with an initial 0.25mm edge crack loaded in mode I at a strain rate of $2 \times 10^3$ $s^{-1}$. Results are shown for the initially rigid cohesive law (left) and the initially elastic cohesive law (right). The gray scale denotes the local yy-stress. The crack does not undergo any macroscopic branching in the initially rigid case while in the initially elastic case the crack bifurcates and travels off axis by many times the cohesive zone length.

**References**
[1] Ravi-Chandar K., Knauss W.G., Int. J. of Fracture, 26(1984) 65; Int. J. of Fracture, 26(1984) 141.
[2] Fineberg J., Gross S.P., Marder M., Swinney H.L., Phys. Rev. Lett., 67(1991) 457;Phys. Rev. B, 45(1992) 5146.
[3] Sharon E., Gross S.P., Fineberg J., Phys. Rev. Lett., 74(1995) 5096; 76(1996) 2117; Phys. Rev. B, 54(1996) 7128.
[4] Sharon E., Fineberg J., Nature, 397(1999) 333.
[5] Xu X.-P., Needleman A., J. Mech. Phys. Solids, 42(1994) 1397.
[6] Camacho G.T., Ortiz M., Int. J. Solids Structures, 33(1996) 2899.
[7] Dugdale D.S., J. Mech. Phys. Solids, 8(1960) 100.
[8] Barenblatt G.I., Adv. Appl. Mech., 7(1962) 56.
[9] Marder M., Gross S., J. Mech. Phys. Solids, 43(1995) 1.
[10] Morrissey J.W., Rice J.R., J. Mech. Phys. Solids, 46(1998) 467; Rice J.R., in Physics of the Earth's Interior (Proc. Int'l School of Physics 'Enrico Fermi,' ed. A.M. Dziewonski and E. Boschi), North-Holland, (1980) 555.
[11] Gao H., J. Mech. Phys. Solids, 44(1996) 1453; Phil. Mag. Lett., 76(1997) 307.
[12] Pandolfi A., Yu C., private communications (2000).